\documentclass[aps,prl,twocolumn,floats,showpacs,amsmath,amssymb,floatfix,%
superscriptaddress]{revtex4}
\usepackage[dvips]{color}
\usepackage{graphicx}
\usepackage{epsfig}
\usepackage{dcolumn}
\usepackage{bm}
\usepackage{amsmath}
\usepackage{amssymb}
\begin{document}
\title{Conversion of charge input into Kondo response}
\author{M.N. Kiselev}
\affiliation{International Centre for Theoretical Physics, Strada
Costiera 11, 34014 Trieste, Italy}
\author{K. Kikoin}
\affiliation{School of Physics and Astronomy, Tel-Aviv University,
Tel-Aviv 69978, Israel}
\author{J. Richert}
\affiliation{Laboratoire de Physique Th\'eorique, Universit\'e
Louis Pasteur and CNRS, 67084 Strasbourg Cedex, France}
\date{\today}

\begin{abstract}
We show how the charge input signal applied to the gate electrode
in double quantum dot may be converted to a pulse in the Kondo
cotunneling current as a spin response of a nano-device in a
strong Coulomb blockade regime. The stochastic component of the input
signal comes as the infrared cutoff of Kondo transmission.
\end{abstract}
\maketitle

Current interest in charge-spin conversion effects is spurred by
challenging prospects of spintronics. Most mechanisms of such a
conversion are related to the interconnection between electrical and
spin current due to spin-orbit interaction (SOI) \cite{Dyak71},
which results in spin accumulation near the sample edges. Such an
accumulation in three- and two-dimensional electron gas in elemental
and III-V semiconductors may result in spin-Hall effect \cite{Munaz}
and positive magnetoresistance \cite{Dyak07}. It was argued also
that the Rashba-type SOI in a quantum dot  induces spin current by a
modulation of the voltages applied to the leads in a three-terminal
device \cite{Lug}. A spin Coulomb drag effect should be mentioned in
this context, which results in spin polarization of the charge
current due to intrinsic friction between electrons with different
spin projections induced by Coulomb scattering
\cite{Cacca76,Davi01}.

In all these propositions the possibilities of {\it conversion} of
charge current into spin current were discussed. It is possible
also to try to use the external electric field for the {\it
generation} of spin current or another spin response. One such
idea was formulated recently for light emitting diodes (LED) based
on conjugated polymers \cite{Li}, where dissociation of excitons
in a strong enough electric field may result in the accumulation of up
and down spin densities near the two ends of the LED.

\begin{figure}[h]
\includegraphics[width=7.0cm,angle=0]{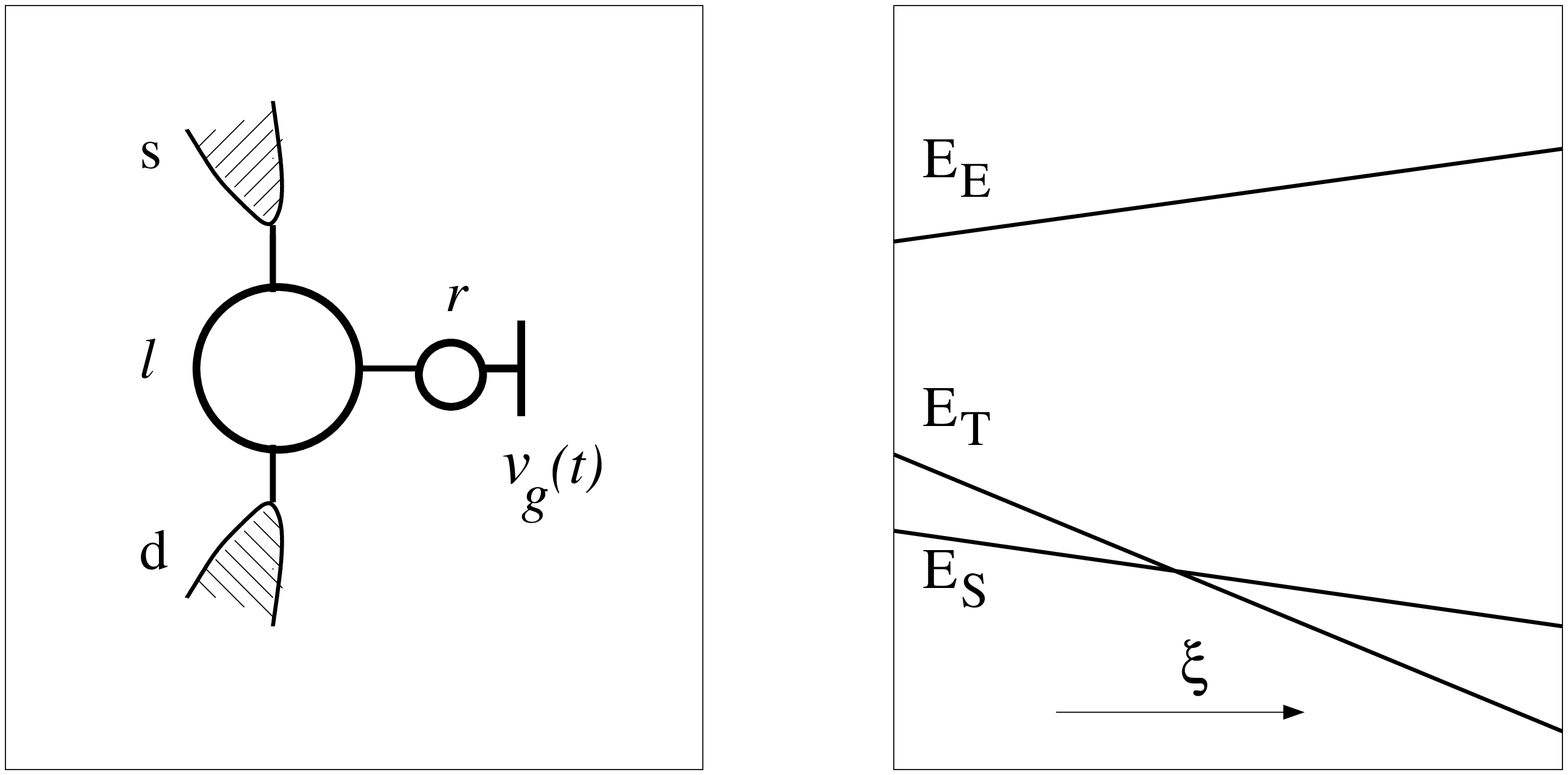}
\caption{Left panel: DQD in a T-shape two-terminal geometry; right
panel: evolution of energy levels  in DQD as a function of scaling
parameter $\xi=\ln D_0/D$ (see text). } \label{f.1}
\end{figure}

In this paper we show that the charge input signal applied to the
gate electrode in a double quantum dot (DQD) may be converted into a
pulse in the Kondo cotunneling current, which is in fact the spin
response of DQD under strong Coulomb blockade.
 We consider the mechanism of activation of internal spin degrees
of freedom by means of a time-dependent potential applied to an
additional electrode to DQD in a contact with metallic leads in
the so called T-shape geometry (Fig. \ref{f.1}).
The electrons in this device are described by the Hamiltonian
\begin{eqnarray}\label{1.0}
&&H=\sum_{j=l,r;\sigma}\left(\varepsilon_j n_j +Q_j n_j^2\right)
+V\sum_{\sigma}(d_{l\sigma}^\dagger d_{r\sigma} + H.c.) \\
&& +  \sum_{b=s,d}\varepsilon^{}_{kb} n^{}_{kb\sigma}+
W\sum_{k\sigma}( c^\dagger_{k\sigma}d_{l\sigma}+ {\rm
h.c.})+V_g(t)n_r \nonumber
\end{eqnarray}
where the three first terms represent electrons in the DQD, electrons
in the leads, and the lead-dot tunneling. The last time-dependent
term stands for the gate voltage applied to the dot $r$. Here $
n_{k\sigma}=c^\dag_{kb\sigma}c^{}_{kb\sigma}$ is the occupation
number for band electrons with the wave vector $k$ and spin
$\sigma$ in the source and drain lead,
$n_{j\sigma}=d^\dag_{j\sigma}d^{}_{j\sigma}$ the electron
number in the left and right dot, and $Q_{(l,r)}$ Coulomb
blockade parameters. The tunneling Hamiltonian involves only
electrons in the left dot. Only the the even standing wave
$c_{k\sigma}=(c_{ks\sigma }+c_{kd\sigma})/\sqrt{2}$ enters the
tunneling Hamiltonian.

The time independent part $H_{dot}^{(0)}$ of the dot Hamiltonian
may be diagonalized in the two-electron charge sector ${\cal
N}=2$, $H_{dot}^{(0)}=\sum_{\Lambda} E_\Lambda
|\Lambda\rangle\langle\Lambda|$. The three lowest states  at
$V_g=0$ are spin triplets $|T\nu\rangle$ with spin projections
$\nu=0,\pm1$, a singlet $|S\rangle$ and a charge transfer singlet
exciton $|E\rangle$ with energies \cite{KA01}
\begin{eqnarray}\label{levels}
E_T =\varepsilon_l + \varepsilon_r, E_S =E_T -2\beta V, E_E
=2\varepsilon_l +Q_l + 2\beta V,
\end{eqnarray}
where $\beta=V/\Delta_{ES}$ and $\varepsilon_{(l,r)}$ are the energy
levels. The static part $V_g(0)$ of the gate
voltage is incorporated in $\varepsilon_r.$ Equations
(\ref{levels}) are obtained for $Q_r\gg Q_l,~~\beta\ll
1$. The ground state of an isolated DQD is a spin singlet $E_S$.

We study the influence of the gate voltage containing both
coherent and stochastic components
\begin{eqnarray}\label{1.3o}
V_g(t)- V_g(0)= v_g(t) = \tilde v_g(t) + \delta v_g(t)
\end{eqnarray}
on the current through DQD. Here $\tilde v_g(t)=\tilde v_g\cos
\Omega t$ is a coherent (deterministic) contribution and $\delta
v_g(t)$ a noise component which is determined by its moments
\begin{eqnarray}\label{1.3}
\overline{\delta v_g(t)}=0,  ~~\overline{\delta v_g(t)\delta
v_g(t')} = \overline{v^2}f(t-t')
\end{eqnarray}
The overline stands for the ensemble average and the
characteristic function $f(t-t')$ will be specified below. We
incorporate $V_g(t)$ into the energy levels (\ref{levels}) by
means of a canonical transformation \cite{KKAR,KNG}
\begin{eqnarray}\label{1.4}
\widetilde{H}_{\rm dot}=U_1H_{\rm dot}U_1^{-1}-i\hbar
\frac{\partial U_1}{\partial t}U_1^\dag,
\end{eqnarray}
with $U_1=\exp[-i\Phi_1(t)n_r]$ and the phase $\Phi_1(t)$ given by
\begin{eqnarray}\label{u1}
\Phi_1(t)=\frac{1}{\hbar} \int^tdt^\prime v_g(t^\prime).
\end{eqnarray}
As a result, in the lowest orders in $V_g$, the time dependent
part of the dot Hamiltonian acquires the form
\begin{eqnarray}\label{ttun}
\delta H_{\rm dot}(t)=-V\left(i\tilde\Phi_1(t){\cal S}_{lr}^{(1)}+
\frac{1}{2}\overline{\Phi_1(t)^2}{\cal S}_{lr}^{(2)}\right)
\end{eqnarray}
where $ {\cal S}_{lr}^{(p)}= \sum_{\sigma}[d^\dagger_{r
\sigma}d_{l\sigma}+(-1)^pd^\dagger_{l \sigma}d_{r\sigma}]$,
\begin{eqnarray}\label{1.5c}
\overline{\Phi_1(t)^2}= \frac{\overline{v^2}}{\hbar^2}\int^t
dt^\prime\int^t dt^{\prime\prime}f(t^\prime-t^{\prime\prime}).
\end{eqnarray}
Thus, the time-dependent gate voltage induces coherent and
stochastic interdot tunneling in DQD described by two terms in
(\ref{ttun}). One may formally include this term in the definition
of the levels (\ref{levels}) and rewrite the dot Hamiltonian as
$H_{dot}{(t)}=\sum_{\Lambda} E_\Lambda(t) X^{\Lambda\Lambda}$
where the level $E_T$ remains intact because the charge
fluctuations do not influence spin degrees of freedom, whereas
\begin{eqnarray}
\label{stadia}
E_S(t) &=& E_S- \delta_{\rm ad}(t)-\delta_{{\rm st},S}(t)  \nonumber \\
E_E(t)&=& E_E + \delta_{\rm ad}(t)+ \delta_{{\rm st},E}(t).
\end{eqnarray}
where the coherent part of renormalization $\delta_{{\rm ad}}(t)
=({2V^2}/\Delta_{ES})\Phi_1^2(t)$ may be considered adiabatically
under the condition $\delta_{\rm ad}/\Delta_{ES}\ll 1$, whereas
the stochastic correction $\delta_{{\rm st},\Lambda}(t)
=({V^2}/4\Delta_{\Lambda T})\overline{\Phi_1^2(t)}$, needs a more
refined treatment. Here $\Delta_{\Lambda\Lambda'}
=|E_\Lambda-E_\Lambda'|$.

Coherent and stochastic components appear also in the cotunneling
part of the effective Hamiltonian $H_{\rm cot}$ which may be
derived by means of the time-dependent Schrieffer-Wolff (SW)
transformation \cite{KNG,KKAR}. Unlike in the standard case of spin
1/2 quantum dots \cite{KNG}, the SW transformation applied to DQD with
the spectrum (\ref{stadia}) intermixes the states
$|\Lambda\rangle$. This intermixing  is described by the operators
$|\Lambda\rangle\langle\Lambda'|$, which form together with
diagonal operators$|\Lambda\rangle\langle\Lambda|$ the set of
generators of the $SO(5)$ group. Ten generators are organized in
three vectors $\bf S$, $\bf P$, $\bf M$ and one scalar $A$. Here
$\bf S$ is the usual S=1 spin operator, $\bf P$ and $\bf M$ are
the vectors describing transitions between spin triplet
$|T\mu\rangle$ and two singlets $|S\rangle$ and $|E\rangle$,
respectively, $A=-i{\cal S}^{(1)}_{lr}/\sqrt{2}$ intermixes the
latter states under the constraint imposed by the Casimir operator
\begin{eqnarray}\label{constr}
{\cal C} = {\bf S}^2 + {\bf P}^2 + {\bf M}^2 + A^2 =4
\end{eqnarray}
(see \cite{KKAR} for further details). The Hamiltonian $H(t)=
H_{\rm dot}(t) + H_{\rm cot}(t)$ may be rewritten in terms of
the above group generators:
\begin{eqnarray}
H_{\rm dot}(t)&=& \frac{1}{2}\left(E_T{\bf S}^2 + \widetilde
E_S{\bf P}^2 + \widetilde E_E{\bf
M}^2 \right) -\mu({\cal C}-4) \nonumber \\
H_{\rm cot}(t)& =& J^{T}_0 {\bf S}\cdot {\bf s} + \widetilde
J^{S}{\bf P}\cdot {\bf s}+\widetilde J^{E}{\bf M}\cdot {\bf
s}.\label{1.7}
\end{eqnarray}
The $SO(5)$ symmetry is preserved by the last term in $H_{\rm
dot}$ by means of the Lagrange multiplier $\mu$.
Tilted coupling parameters are time-dependent. The charge-spin
conversion mechanism under discussion is a manifestation of the
$SO(5)$ dynamical symmetry of DQD.

The Hamiltonian (\ref{1.7}) describes Kondo cotunneling through
DQD in presence of time-dependent perturbations (\ref{1.3o}) Its
coherent component is responsible for the conversion of the charge
signal $\tilde v_g(t)$ into a Kondo-type zero bias anomaly in
tunnel current response, which arises in the spin channel. Its
stochastic component $\delta v_g(t)$ results in the correction of
charge noise into incoherent corrections to Kondo cotunneling.

\begin{figure}[h]
\includegraphics[width=4cm,angle=0]{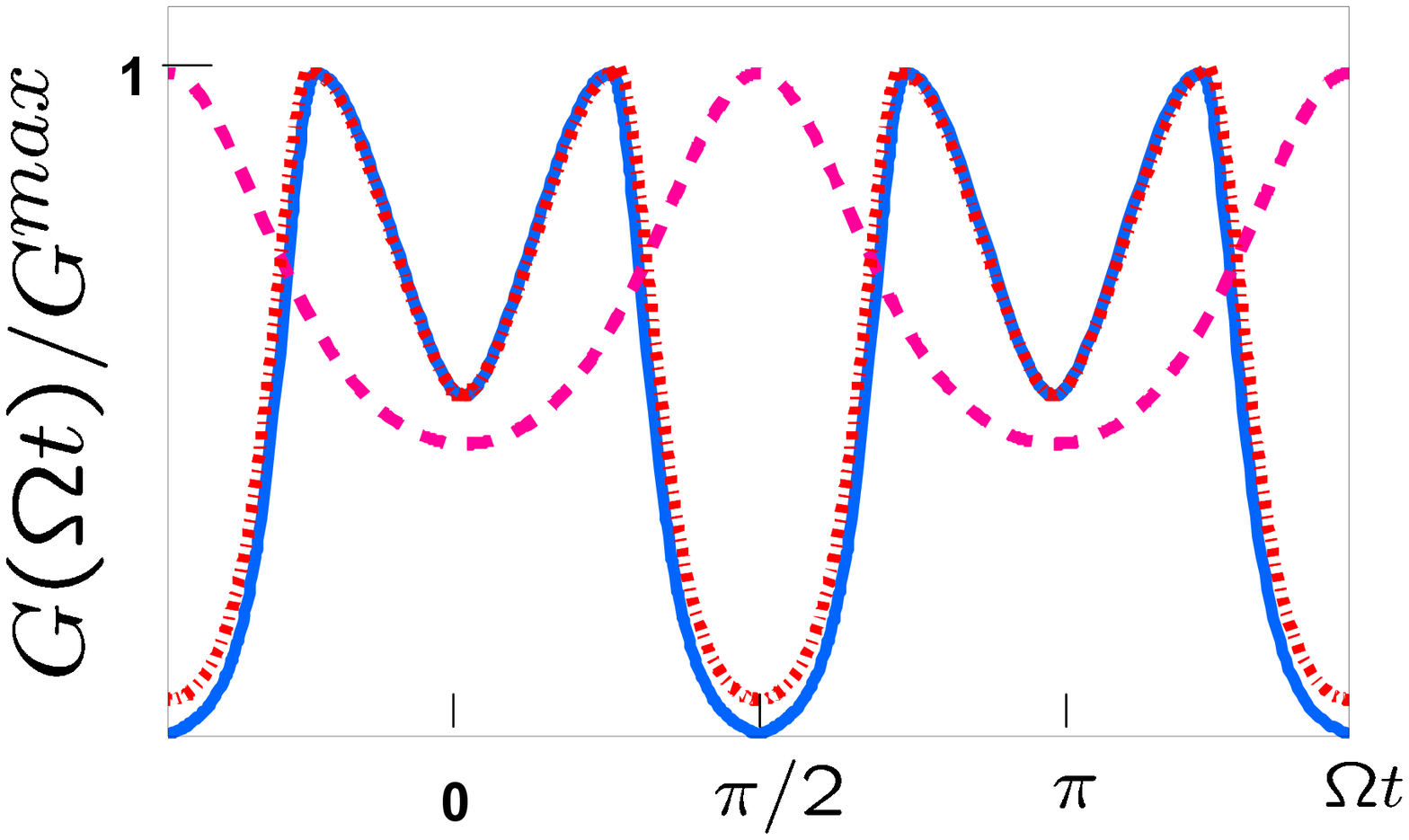}\hspace*{5mm}
\includegraphics[width=4cm,angle=0]{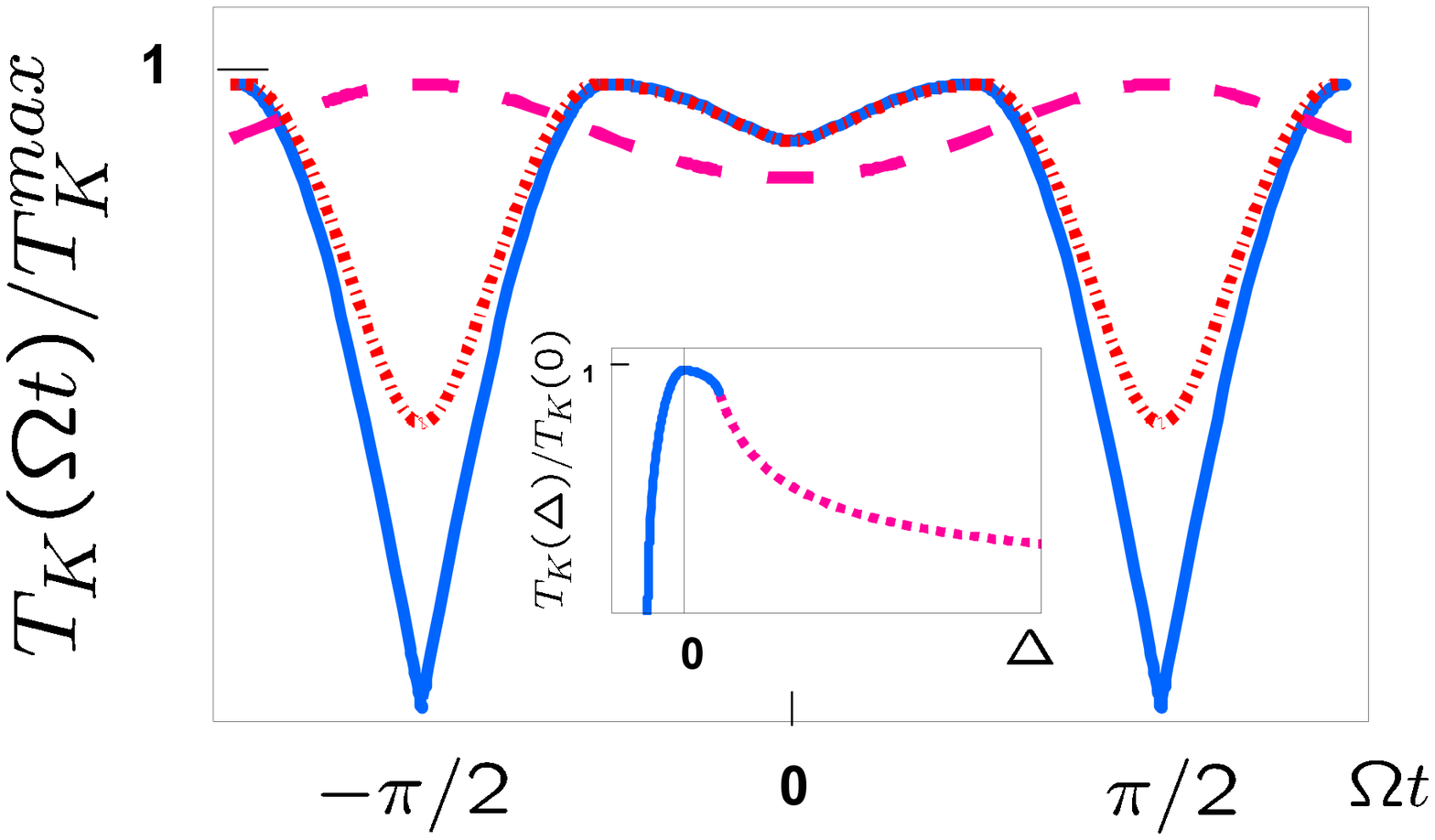}
\caption{(Color online) Left panel: Oscillations of ZBA in tunnel
conductance due to time-dependent gate voltage $\tilde v_g\cos
\Omega t$ corresponding to the ST-gap oscillating around large
positive $\Delta_{ST}^0\gg T_{K0}$  (dashed line), around zero
$\Delta$ down to $T^{min}_K=T_{K0}/2$ (dash-dotted line) and down to
$T^{min}_K=0$ (solid line), respectively. Right panel: time
dependent $T_K$ (notations are the same as for the left panel);
insert: $T_K$ as a function of $\Delta_{ST}$.  Calculations are
performed at $T/T_{K0}=1.4 , \eta=0.5.$}\label{f.2}
\end{figure}

The coherent part of the time-dependent Kondo problem can be solved in
adiabatic approximation \cite{KKAR,KNG}. As a result of a time
dependent Schrieffer-Wolff transformation and elimination of
high-energy states, additional renormalizations $M_\Lambda(\xi)$ of
the levels $E_\Lambda$  arise where $\xi=\ln (D_0/D)$ is the
scaling variable \cite{KA01,Hald} Then the singlet-triplet gap
transforms into
 \begin{eqnarray}
\Delta_{ST}(t) = \Delta_{ST}^0+M_T - M_S  -\delta_{\rm ad} (t).
 \label{1.8}
\end{eqnarray}
and the singlet-triplet crossover (Fig. \ref{f.1}, right panel)
takes place, provided the self-consistent condition
$T_K(\Delta_{ST},t)>\Delta_{ST}(t)$ is satisfied
\cite{KA01,Hald,Kogan,Eto,Pust1}. $T_K$ is a sharp function of
$\Delta_{ST}$ with a maximum $T_{K0}$ at $\Delta_{ST}=0$ (inset in
the left panel of Fig. \ref{f.2}). Its right slope is described by
the ratio
\begin{equation}
\label{asymp}
    \frac{T_K(\Delta_{ST},t)}{T_{K0}}=\left[\frac{T_{K0}}{\Delta_{ST}(t)}
    \right]^\eta,
    \end{equation}
valid for intermediate asymptotic positive values of $\Delta_{ST}$
at $T_{K0}/\Delta_{ST}\lesssim 1$ (dotted part of the curve
$T_K(\Delta)/T_K(0)$ in the inset). Here $\eta<1$ is a universal
constant.
This sharp dependence is a key to the transformation mechanism of
charge input into Kondo conductance response.

We estimate the influence of $\tilde v_g(t)$ on the tunnel
conductance $G(T,t)$ at given $T > T_K$ in a situation where the
adiabatic temporal variations of $T_K(\Delta_{ST},t)$ take place.
Then the tunnel conductance obeys the law
\begin{eqnarray}\label{conduct}
G/G_0\sim \ln^{-2}(T/T_K).
\end{eqnarray}
Substituting (\ref{asymp}) in (\ref{conduct}), one gets
\begin{eqnarray}\label{conducttt}
{G(t)}/{G_0} \sim\left(\ln(T/T_{K0})-\eta\ln
(T_{K0}/\Delta_{ST}(t) \right)^{-2}
\end{eqnarray}
The curves $G(\Omega t)$ shown in Fig. \ref{f.2} (left panel)
describe transformations of an oscillating signal in the charge
perturbation $\tilde v_g(t)$ into oscillations of Kondo-type zero
bias anomaly (ZBA) of tunnel conductance. The transformation effect
is especially distinct provided the oscillations of $v_g(t)$ change
the sign of $\Delta_{ST}$, i.e. induce an $S\to T$ crossover (solid
and dash-dotted curves in Fig. \ref{f.2} corresponding to the solid
part of $T_K(\Delta)$ curve in the inset). It is worthwhile to
notice that this mechanism of the adiabatic transformation of a
charge signal into Kondo response is close to the one  proposed for
Kondo shuttling \cite{shut} where the source of time-dependence are
the nanoelectromechanical oscillations of a quantum dot.

The mechanism which converts a stochastic component $\delta v_g(t)$
of the input signal into a stochastic spin response is quite
unusual. Instead of dephasing due to time-dependent spin flip
processes \cite{KNG}, stochastization of the energy spectrum of DQD
results in the loss of a Curie-type spin response at some
characteristic energy $\zeta$. This effect is related to the time
dependence of the factor $\tilde \mu(t)$ in the Hamiltonian
(\ref{1.7}). Indeed, inserting (\ref{stadia}) into (\ref{1.7}), one
may write the stochastic part of $H_{\rm dot}$ as
\begin{eqnarray}\label{cost}
H_{\rm dot}^{\rm st}= [\delta_{{\rm st},S}(t){\bf P}^2 -
\delta_{{\rm st},E}(t){\bf M}^2]/2
\end{eqnarray}
Unlike the adiabatic part of time dependent energy levels $E_\Lambda(t)$
incorporated in (\ref{1.8}), this term describes fluctuations due
to the dynamical symmetry of DQD. At energies $\sim T_K\ll E_E$ the
state $|E\rangle$ is frozen out. Then instead of the exact Casimir
constraint (\ref{constr}) for the $SO(5)$ group, one deals with a
fluctuating constraint ${\cal \tilde C}={\bf S}^2 +{\bf P}^2$ for
the reduced group $SO(4)$ describing an ST multiplet where the
fluctuating part may be written in the form $\mu_{\rm st}(t){\bf
P}^2=-\mu_{\rm st}(t){\bf S}^2$ with $\mu_{\rm st}(t)=\delta_{{\rm
st},S}(t)/2$. At $T\to 0$ where the singlet is also frozen out, one
arrives to the effective dot Hamiltonian
\begin{eqnarray}\label{constrac}
H_{\rm dot}(t)&=& \frac{1}{2}E_T{\bf S}^2  -\mu_{\rm ad}({\bf S}^2
-2)
-\mu_{{\rm st},S}(t){\bf S}^2~\nonumber \\
&=&\sum_{\nu=0,\pm 1} [\varepsilon - \mu(t)]f^{\dag}_\nu f^{}_\nu .
\end{eqnarray}
where the fermion representation for S=1, $|T\nu\rangle\langle
T{\nu'}|= f^\dag_\nu f^{}_{\nu'}$, is used in the second line.
 Here $\varepsilon=E_T/2$, and the time-dependent chemical
potential for spin fermions is defined as $\mu(t)= \mu_{\rm ad} -
\mu_{{\rm st},S}(t)$. The stochastic component of $\mu$ may be
treated as a random potential in the time domain which describes
the fluctuations of the global fermionic constraint \cite{KKAR}.
Then the propagation of spin fermions in the random time-dependent
potential $\mu_{{\rm st},S}(t)$ may be studied by means of the
"cross technique" developed for the calculation of electron propagation
in a field of impurities randomly distributed in real space.

The frequency $\Omega$ in $\tilde v_g(t)$ is the slowest frequency
in our problem, and $\mu_{{\rm st},S}(t)$ also varies slowly in
time, so that the relaxation time $\tau = \hbar/\gamma$ in the
noise correlation function
$D(t-t')=\hbar^2\langle\mu(t)\mu(t')\rangle \sim
\exp[-\gamma(t-t')]$ is a longest time in the model. Then one may
take the limit
\begin{eqnarray}
D(\omega)=\lim_{\gamma\to 0}
\frac{2\zeta^2\gamma}{\omega^2+\gamma^2}=2\pi \zeta^2\delta(\omega)
\label{fluct}
\end{eqnarray}
for its Fourier transform, $\zeta$ being the bandwidth of the
Gaussian correlation. In this limit the averaged spin propagator
describes the ensemble of states with chemical potential $\mu_{\rm
st}$$=const$ in a given state, but this constant is random in each
realization \cite{foot}. Thus the problem of decoherence of the spin
state in stochastically perturbed DQD is mapped on the so-called
Keldysh model \cite{keld65,efros70,sad} originally formulated for
$\delta$-correlated impurity scattering potentials in momentum
space. The problem can be solved exactly and the decoherence time is
fixed by $\zeta$ of the Gaussian.
\begin{figure}[h]
  \includegraphics[width=1.5cm,angle=0]{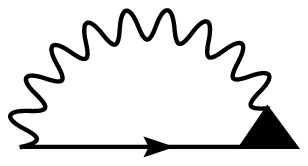}\hspace*{8mm}
  \includegraphics[width=1.2cm,angle=0]{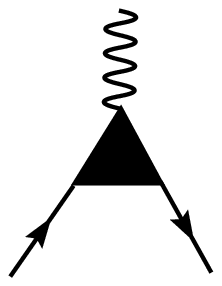}\hspace*{8mm}
  \includegraphics[width=1.5cm,angle=0]{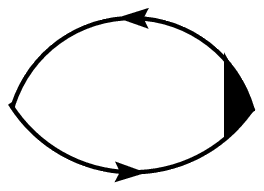}\hspace*{8mm}
  \includegraphics[width=1.5cm,angle=0]{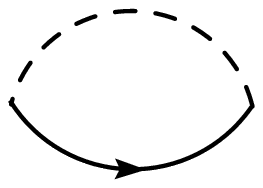}
  \caption{ Feynmann diagrams for the self energy $\Sigma$
  with vertex corrections, triangular vertex $\Gamma$, spin susceptibility $\chi$
   and parquet Kondo loops $K$ (from left to right). Solid, dashed and wavy lines
  denote  bare spin-fermion propagator $G$, conduction electron propagator
  $g$ and correlation function $D$.}\label{f.diag}
\end{figure}

The solution of the Dyson equation obeyed by
$G^{-1}(\epsilon)=\epsilon +\mu_0-\Sigma(\epsilon)$ (Fig.
\ref{f.diag}) \cite{com2}, the Fourier transform of the spin-fermion
propagator $ G^R_{T\nu}(t-t')=\langle
f^{}_\nu(t)f^\dag_\nu(t')\rangle_R=-i\langle[f^{}_\nu(t)f^\dag_\nu(t')]_+\rangle
$, is given by
\begin{eqnarray}\label{GF1}
\Sigma(\epsilon)=\int\frac{d\omega}{2\pi}
\Gamma(\epsilon,\epsilon-\omega;\omega)G(\epsilon-\omega)D(\omega)
=\zeta^2\Gamma(\epsilon,\epsilon;0)G(\epsilon)\nonumber
\end{eqnarray}
(the index $\nu$ is omitted, since the fluctuations of $\mu$ are
related to the global $U(1)$ symmetry). Using the Ward identity for
a triangular vertex shown in Fig. \ref{f.diag},
    $\Gamma(\epsilon,\epsilon;0)=dG^{-1}(\epsilon)/d \epsilon$, we
transform the Dyson equation into differential equation
\begin{eqnarray}
\zeta^2 d G/d x+ xG -1=0 \label{DE}
\end{eqnarray}
where $x=\epsilon+\mu_0$ with a solution \cite{efros70}
\begin{eqnarray}
G^R(x)=\frac{1}{\zeta\sqrt{2\pi}}
\int_{-\infty}^{\infty}e^{-z^2/2\zeta^2}\frac{dz}{x-z+i\delta}
\label{GR}
\end{eqnarray}
representing the set of spin states with stochastic chemical
potential averaged with a gaussian exponent characterized by the
dispersion $\zeta$. Remarkably, $G^R(x)$ has no poles,
singularities or branch cuts.

To check the spin properties of stochasticized DQD, we calculate
its spin response at finite $T$ determined as
    $\chi(i\omega_m)= T\sum_n{\cal
G}(i\omega_m+i\epsilon_n){\cal
G}(i\epsilon_n)\Gamma(i\epsilon_n,i\epsilon_n+i\omega_m;i\omega_m)
$ where ${\cal G}(i\epsilon_n)$ is the Matsubara continuation of
(\ref{GR}) on to the imaginary axis (see Fig. \ref{f.diag}).
 The same Ward identity provides the exact equation
for the vertex
\begin{eqnarray}
\Gamma^R(\epsilon,\epsilon;0)=(\epsilon G^R-1)/\zeta^2(G^R)^2,
\label{GamR}
\end{eqnarray}
which allows to calculate the static susceptibility
\begin{eqnarray}
\chi(\omega=0,T)=\frac{1}{\sqrt{8\pi}\zeta}\int_{-\infty}^{\infty}dx
e^{-x^2/2} x\tanh\left(\frac{x\zeta}{2T}\right) \label{ssuca}
\end{eqnarray}
From its temperature dependence plotted in Fig. (4) we see that the
the Curie-type spin response $\chi(0,T)\sim 1/T$ at high $T\gg
\zeta$ transforms into a constant at zero $T$, $\chi(0,0) \sim
1/\zeta$.  This means that the DQD looses at $\{\omega,T\}\ll \zeta$
the characteristics of a localized spin due stochastization, so it
cannot serve as a source of Kondo screening at low energies. The
frequency dependence of ${{\rm Im} \chi}(\omega,T)$ (Fig.4, inset)
confirms this conclusion. \vspace*{-2mm}
\begin{figure}[h]
\includegraphics[width=4cm,angle=0]{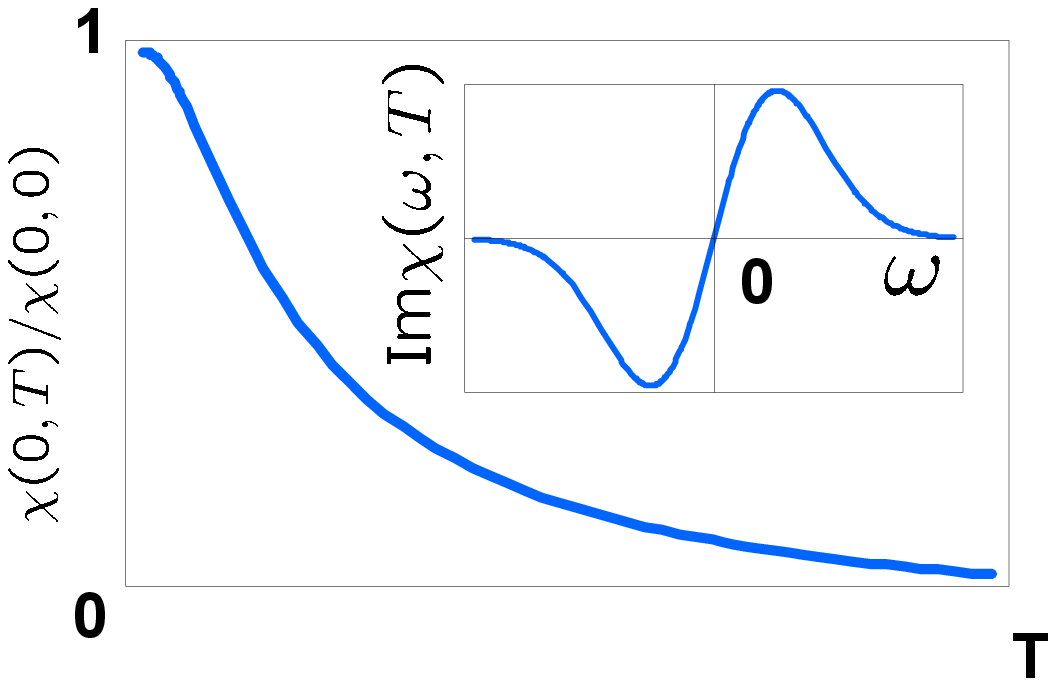}\vspace*{-3mm}
\caption{(Color online) Static susceptibility $\chi(0,T)$. Inset
shows a frequency dependence of ${\rm Im}\chi$ with a maximum at
$\omega \sim \zeta$.}\label{f.4}
\end{figure}\vspace*{-2mm}
Such a behavior of the DQD affects the Kondo response described by
$H_{\rm cot}$ (\ref{1.7}). Time dependence in the vertex $\Gamma$
(Fig. \ref{f.diag}) results in dephasing of Kondo cotunneling
\cite{KKAR}, which, however, is exponentially weak at $T\ll
\Delta_{ST}$. More significant is the influence of stochastization
of the DQD. Inserting (\ref{GR}) into the Kondo loop $K
(i\epsilon_n) \sim J^2 T\sum_m \int \frac{d {\bf p}}{(2\pi)^3}g({\bf
p},i\omega_m){\cal G}(i\epsilon_n\pm i\omega_m)$ responsible for
logarithmic singularity in conventional Kondo scattering (Fig.
\ref{f.diag}), one obtains a combination of logarithmic,
hypergeometric and imaginary error functions \vspace*{-1mm}
\begin{eqnarray}
K (\epsilon\to 0)/J=\rho_0 J\ln(\sqrt{2C}D/\zeta)+ \nonumber
\end{eqnarray}
\vspace*{-7mm}
\begin{eqnarray}
+\frac{1}{2}\rho_0
J\left[_1F_1\left(\frac{3}{2},2,\frac{T^2}{2\zeta^2}\right)\left(\frac{T}{\zeta}\right)^2
-\pi {\tt Erfi}\left(\frac{T}{\sqrt{2}\zeta}\right)\right]
\nonumber\end{eqnarray} where $\gamma=\ln C$ is the Euler
constant. In two limiting cases of low and high temperatures
relative to the dispersion $\zeta$ of the noise spectrum, it leads to
following expressions \vspace*{-3mm}
\begin{eqnarray}
K(\epsilon\to 0)/J=\left\{
\begin{array}{c}
\rho_0 J\ln(D/T),\;\;\; T\gg \zeta\\
\\\label{Gm}
\rho_0 J\ln(D/\zeta),\;\;\; \zeta\gg T\\
\end{array}\right.
\end{eqnarray}
Thus the noise amplitude plays the role of the infrared cut-off
(similarly to Kondo-spin glass problem \cite{kisop00}). This
cut-off distorts the coherent pulses in ZBA (Fig. \ref{f.2})
provided $\zeta$ is comparable with $T_K$ .

In conclusion, we presented and discussed the conversion mechanism
of a time-dependent coherent and stochastic input signal at a gate
electrode into a Kondo-type spin response in tunnel conductance.
This mechanism is related to the dynamical symmetry of DQD.

Authors are indebted to F.Marquardt, N.Prokof'ev, M.V. Sadovskii and
T. Ziman for valuable advices.


\begin{thebibliography}{99}
\bibitem{foot} The analog of this ensemble in real space is the
inhomogeneous broadening of a ESR signal.
\bibitem{com2} Here $\mu_0$ is a constant non-random part
of $\mu(t)$.
\end{thebibliography}
\end{document}